\documentclass[aps,pra,twocolumn,nofootinbib,floatfix,10pt]{revtex4-2}

\usepackage{amsmath}
\usepackage{amssymb}
\usepackage{wasysym}
\usepackage{graphicx}
\usepackage{color,soul}
\usepackage{physics}
\usepackage{siunitx}
\usepackage{dsfont}
\usepackage{float}
\usepackage{mathtools}
\usepackage[hidelinks,colorlinks=true,linkcolor=black,citecolor=black,urlcolor=black]{hyperref}
\usepackage[colorinlistoftodos]{todonotes}

\usepackage{lineno} 
\usepackage{etoolbox} 
\pretocmd
\align{\linenomath}{}{} 
\apptocmd
\endalign{\endlinenomath}{}{} 
\usepackage{etoolbox}
\makeatletter
\patchcmd\linenumberpar{\@LN@parpgbrk}{\penalty\@LN@parpgpen\relax}{}{}
\makeatother
\modulolinenumbers[2]
\usepackage[normalem]{ulem}

\usepackage[english,nomargin,inline,marginclue,draft]{fixme}
\fxusetheme{colorsig}
\FXRegisterAuthor{th}{ath}{\color{blue}TH}  
\FXRegisterAuthor{ib}{aib}{\color{red}IB}
\FXRegisterAuthor{sh}{ash}{\color{cyan}SH}
\FXRegisterAuthor{db}{adb}{\color{blue}DB}
\FXRegisterAuthor{tc}{atc}{TC} 
\makeatletter
\renewcommand*\FXLayoutInline[3]{%
  {\@fxuseface{inline}\ignorespaces{\color{fx#1}[#3: #2]}}}
\makeatother

\long\def\symbolfootnote[#1]#2{\begingroup%
\def\thefootnote{\fnsymbol{footnote}}\footnotetext[#1]{#2}\endgroup}

\def\nobreakbefore{%
  \relax\ifvmode\else
    \ifhmode
      \ifdim\lastskip > 0pt\relax
        \unskip\nobreakspace
      \else 
        \nobreakspace
      \fi
    \fi
  \fi
}
\let\oldcite\cite
\renewcommand\cite{\nobreakbefore\oldcite}

\newcommand{\Jy}{\ensuremath{J_y}}

\newcommand{\gh}{\ensuremath{g_{\mathrm{hh}}^{(2)}}\,}

\newcommand{\kB}{\ensuremath{k_{\rm B}}} 

\newcommand{\ivec}{\ensuremath{\textit{\textbf{i}}}}
\newcommand{\jvec}{\ensuremath{\textit{\textbf{j}}}}
\newcommand{\kvec}{\ensuremath{\textit{\textbf{k}}}}

\newcommand{\dvec}{\ensuremath{\textit{\textbf{d}}}}

\DeclareSIUnit\Gauss{G}
\DeclareSIUnit\rad{rad}
\DeclareSIUnit\mrad{mrad}
\DeclareSIUnit\Erx{E_r^x}
\DeclareSIUnit\Ery{E_r^y}
\DeclareSIUnit\Eryl{E_r^{y,L}}

\begin{document}
	
\title{Formation of stripes in a mixed-dimensional cold-atom Fermi-Hubbard system}

\author{Dominik~Bourgund$^{1,2,\ast}$}
\author{Thomas~Chalopin$^{1,2}$}
\author{Petar~Bojovi\'{c}$^{1,2}$}
\author{Henning~Schl\"omer$^{2,3,4}$}
\author{Si~Wang$^{1,2}$}
\author{Titus~Franz$^{1,2}$}
\author{Sarah~Hirthe$^{1,2,\dagger}$}
\author{Annabelle~Bohrdt$^{2,5}$}
\author{Fabian~Grusdt$^{2,3,4}$}
\author{Immanuel~Bloch$^{1,2,4}$}
\author{Timon~A.~Hilker$^{1,2}$}


\affiliation{$^1$Max-Planck-Institut f\"{u}r Quantenoptik, 85748 Garching, Germany}
\affiliation{$^2$Munich Center for Quantum Science and Technology, 80799 Munich, Germany}
\affiliation{$^3$Arnold Sommerfeld Center for Theoretical Physics (ASC), Ludwig-Maximilians-Universit\"{a}t, 80333 Munich, Germany}
\affiliation{$^4$Fakult\"{a}t f\"{u}r Physik, Ludwig-Maximilians-Universit\"{a}t, 80799 Munich, Germany}
\affiliation{$^5$Institute of Theoretical Physics, University of Regensburg, 93053 Regensburg, Germany}

\symbolfootnote[1]{Electronic address: {dominik.bourgund@mpq.mpg.de,\\ timon.hilker@mpq.mpg.de}}
\symbolfootnote[2]{Present address: ICFO - Institut de Ciencies Fotoniques, The Barcelona Institute of Science and Technology, Castelldefels (Barcelona) 08860, Spain}

 \begin{abstract}
 	The relation between $d$-wave superconductivity and stripes is fundamental to the understanding of ordered phases in cuprates~\cite{Tranquada1995, Axe1994, Fujita2002, Scalapino2012, Kivelson2003, Tranquada2020}.
 	While experimentally both phases are found in close proximity, numerical studies on the related Fermi-Hubbard model have long been investigating whether stripes precede, compete or coexist with superconductivity~\cite{Emery1997, Daou2010,Qin2020}.
 	Such stripes are characterised by interleaved charge and spin density wave ordering where fluctuating lines of dopants separate domains of opposite antiferromagnetic order~\cite{Zaanen1989,Schulz1989,Poilblanc1989}.
 	Here we show first signatures of stripes in a cold-atom Fermi-Hubbard quantum simulator. By engineering a mixed-dimensional system, we increase their typical energy scales to the spin exchange energy, enabling us to access the interesting crossover temperature regime where stripes begin to form~\cite{Schloemer2022}. We observe extended, attractive correlations between hole dopants and find an increased probability to form larger structures akin to stripes. 
 	In the spin sector, we study correlation functions up to third order and find results consistent with stripe formation.
 	These higher-order correlation measurements pave the way towards an improved microscopic understanding of the emergent properties of stripes and their relation to other competing phases. 
 	More generally, our approach has direct relevance for newly discovered high-temperature superconducting materials in which mixed dimensions play an essential role~\cite{Sun2023,Qu2023,Oh2023,Schloemer2023}.
\end{abstract}
\maketitle

\section{Introduction}
Low-temperature phases in cuprate materials have been under intense scrutiny from both experimental and theoretical investigations for over forty years while still eluding full understanding \cite{Kivelson2003,Scalapino2012,Tranquada2020}. The repulsive, two-dimensional (2d), spin-1/2 Fermi-Hubbard model and its natural extensions are widely assumed to provide minimal models to explain the origin of high-temperature superconductivity in materials with insulating behaviour in their undoped state. 
Meanwhile, many experimental studies in solids~\cite{Tranquada1995, Axe1994,Fujita2002,Abbamonte2005,Parker2010} as well as theoretical work~\cite{Zaanen1989,Schulz1989,Poilblanc1989,Machida1994,White1998, Zheng2017,Huang2018,Qin2020,Wietek2021} found stripe phases featuring charge density waves in combination with incommensurate antiferromagnetic (AFM) order in the doped cuprates and Fermi-Hubbard model, competing with superconductivity~\cite{Qin2020}. At elevated temperatures, these ordered phases give way to an even less understood normal phase, which has been argued to feature non-Fermi liquid properties with deconfined spin- or charge excitations~\cite{Chowdhury2015,Zhang2020}.

Ultracold atoms in optical lattices provide natural implementations of the Fermi-Hubbard model with a high degree of control over system parameters~\cite{Esslinger2010}. While solid-state experiments mostly focus on spectroscopic and dynamical response measurements, quantum simulation, especially with single-site resolution, opened up access to new sets of microscopic observables and correlation functions \cite{Parsons2015, Haller2015,Cheuk2015,Omran2015}. Previous studies found AFM correlations \cite{Greif2013, Hart2015,Boll2016,Mazurenko2017, Parsons2016, Cheuk2016}, investigated the effect of doping on the spin order \cite{Koepsell2019,Hartke2020,Koepsell2021,Chiu2019,Ji2021,Bohrdt2021,Hartke2023} and observed pairing of dopants in tailored ladder systems \cite{Hirthe2023}. 

Here, we present the first observation of hole attraction beyond nearest-neighbouring sites and signatures of stripes in a repulsive, 2d Hubbard system with mixed-dimensional coupling using charge and spin correlators. We report on collective behaviour of multiple dopants directly from real-space snapshots using higher-order correlation functions. 

\section{Experimental implementation}

\begin{figure*}[t]
	\centering
	\includegraphics[scale = 1]{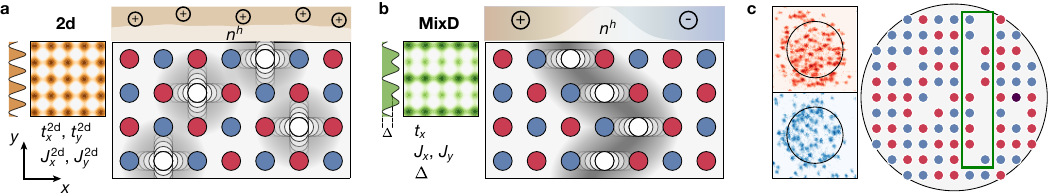}
	\caption{\textbf{Mixed-dimensional Fermi-Hubbard systems.} \textbf{a}, Illustration of the isotropic 2d Fermi-Hubbard model. Holes delocalise within small regions and disturb their respective spin background, forming magnetic polarons. The overall hole density is uniform and holes repel each other due to their fermionic statistics at experimentally accessible temperatures of $\kB T\approx J$. There are no domain walls in the spin order. \textbf{b}, By raising the potential on every other lattice site along $y$ by $\Delta$, we suppress tunnelling along this direction, thus removing the Pauli repulsion between holes, while preserving the superexchange coupling \Jy. The holes form collective structures, which also result in a domain wall in the AFM correlations of the system, indicated by the AFM parity change across the stripe. \textbf{c}, A single raw experimental shot of spin-up (red), spin-down (blue) atoms and doubly occupied sites (purple) as well as its reconstructed spin and charge distribution with the main system being inside the black circle, surrounded by a low density reservoir (see \hyperref[sec:sequence]{SI}). The green box indicates a stripe-like structure.}
	\label{fig:fig1}
\end{figure*}
The isotropic Fermi-Hubbard model is governed by the competition between the kinetic energy of dopants favouring delocalisation and the magnetic energy of the AFM spin order which is disrupted by dopant motion. 
Consequently, the energy scale at which stripe order is expected to occur is only a few percent of the tunnelling energy~\cite{Wietek2021}, placing it out of reach for state-of-the-art cold-atom quantum simulators.
In particular, for temperatures around the superexchange energy, the effective repulsion due to the fermionic nature of the holes (Pauli blocking) disfavours tightly-bound hole pairs and extended structures like stripes while favouring the formation of magnetic polarons (see Fig.~\ref{fig:fig1}a)~\cite{Koepsell2019}.

We tilt the balance in favour of collective charge and spin ordering by restricting the hole motion to one dimension (1d), thus reducing the kinetic energy while keeping spin couplings two-dimensional. 
This leads to an increase in the characteristic energy scales of collective effects to experimentally accessible regimes as kinetic and magnetic terms in the Hamiltonian are less frustrated~\cite{Bohrdt2022, Hirthe2023, Schloemer2022}. 
In this mixed-dimensional (mixD) setting, we thereby bias hole attraction and stripe formation along the direction perpendicular to the hole motion. Thus we favour fully filled stripes while retaining the key concepts of charge and spin density wave ordering associated with the stripe phase (see Fig.~\ref{fig:fig1}b) \cite{Huang2018, Wietek2021}. Here we engineer a mixD setting by applying a potential offset to every other chain within the lattice. For sufficiently large offsets, this removes nearest-neighbour hopping along the perpendicular direction while slightly increasing spin couplings \cite{Duan2003,Trotzky2008}. 

In the experiment, we realise the spin-1/2 Fermi-Hubbard model by using $^6\mathrm{Li}$ atoms in an optical superlattice with a homogeneous, circular system of $\sim110$ sites surrounded by a low-density reservoir (see Fig.~\ref{fig:fig1}c). In the limit of strong on-site interactions $U$, the essential physics of the system can be captured by the $t-J$ Hamiltonian using projections $\mathcal{P}$ onto singly occupied sites,
\begin{multline}\label{eq:tJ}
	\hat{\mathcal{H}}_{t-J} = \sum_{\langle \ivec,\jvec \rangle,\sigma} \hat{\mathcal{P}}\left(-t_{\ivec\jvec} \hat{c}_{\ivec,\sigma}^\dagger \hat{c}_{\jvec,\sigma} + \mathrm{h.c.}\right)\hat{\mathcal{P}} +\\
	+ \sum_{\langle \ivec,\jvec\rangle,\sigma} J_{\ivec\jvec}\left(\hat{\mathbf{S}}_\ivec \cdot \hat{\mathbf{S}}_\jvec - \frac{\hat{n}_\ivec\hat{n}_\jvec}{4}\right),
\end{multline}
with tunnel couplings $t_{\ivec\jvec}\in \{t_x, t_y\}$, spin exchange couplings $J_{\ivec\jvec} \in \{J_x, J_y\}$, spin-up/down fermionic creation (annihilation) operators on site $\ivec$, $\hat{c}_{\ivec,\uparrow/\downarrow}^\dagger$ ($\hat{c}_{\ivec,\uparrow/\downarrow}$), and on-site spin (density)  operators $\hat{\mathbf{S}}_\ivec$ ($\hat{n}_\ivec$). This model suffers from the Fermion sign problem, making it numerically challenging to tackle even in the mixD regime~\cite{Dicke2023}.

Here we work at $U/t_x = 27(2)$, $J_y/t_x = 0.6(2)$ and a filling of $n \approx 0.7-0.9$ (hole doping $\delta = 1-n$) with a temperature of $\kB T/t_x=0.3(1)$ (see \hyperref[sec:sequence]{SI}). We make use of an optical superlattice along $y$ to controllably detune neighbouring sites by $\Delta = 0.65(5)\,U \gg t_x, t_y^{\mathrm{2d}}$, thus effectively disabling nearest-neighbour tunnelling along $y$ ($t_y\approx0$), and leading to a spin coupling $J_y = 2(t_y^{\mathrm{2d}})^2 \left(\frac{1}{U-\Delta}+\frac{1}{U+\Delta}\right)$, where $t_y^{\mathrm{2d}}$ is the tunnel coupling in the 2d system without potential offsets. Due to the staggered superlattice potential, there is also a second-order next-nearest-neighbour hopping term along $y$, which reintroduces a weak Pauli repulsion at distance $d_y=2$. This term, however, is smaller than $J_y$, such that it is still expected to be favourable for stripes to form (see \hyperref[sec:sequence]{SI} for more details on preparation and subdominant couplings).

\section{Hole-hole correlations}

\begin{figure}[t]
	\centering
	\includegraphics[scale = 1]{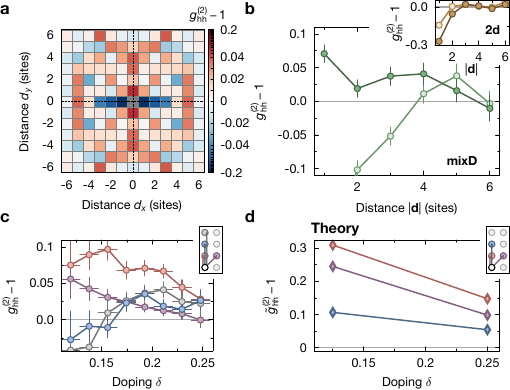}
	\caption{\textbf{Hole correlations beyond nearest neighbours.} \textbf{a}, Hole-hole correlations in a mixD system revealing bunching (attraction) along $y$ at distances $d_y\geq 1$ and antibunching (repulsion) along $x$ with $\delta = 0.18$. The symmetrisation is indicated by the dashed lines. A cut along $y$ ($x$) is shown in dark (light) green for mixD systems in \textbf{b}, with the outset showing the equivalent data for standard 2d systems. The dependency of the mixD correlator $\gh$ at distance $\dvec=(0,1)$, $(1,1)$, $(0,2)$, $(0,3)$ on doping is plotted in \textbf{c} in red, purple, blue, grey with a doping binning of $\pm0.009$. In a doping region around 0.2, the correlators for distances $d_y>1$ are positive, indicating longer range charge correlations. Error bars are estimated using bootstrapping. In \textbf{d} we show results for the renormalised correlator (see \hyperref[sec:numerics]{SI}) from DMRG calculations for a system size of $L_x \times L_y = 8\times 3$ as a function of doping for  $\kB T/t_x = 0.41$}
	\label{fig:fig2}
\end{figure}
In order to reveal the charge order within the system, we evaluate the connected, normalised two-point hole-hole correlator
\begin{equation}\label{eq:hh}
	g^{(2)}_{\mathrm{hh}}(\dvec) -1 = \frac{1}{\mathcal{N_\dvec}}\sum_{\ivec} \left(\frac{\langle \hat{n}^h_\ivec \hat{n}^h_{\ivec+\dvec} \rangle}{\langle \hat{n}^h_\ivec\rangle \langle \hat{n}^h_{\ivec+\dvec} \rangle} - 1 - o_\delta\right),
\end{equation}
with hole density operator $\hat{n}^h_\ivec$ at position \ivec\,, and normalisation $\mathcal{N_\dvec}$. Due to the finite size and particle number fluctuations in our system, there is a global, doping-dependent offset $o_\delta\approx -0.03$ on this correlator that we subtract (see \hyperref[sec:corrs]{SI}). A positive (negative) value of this correlator indicates attraction (repulsion) between holes at distance $\dvec$.

We consider hole correlations in a mixD system with a doping of $\delta = 0.18$ in Fig.~\ref{fig:fig2}a,b. We observe a positive nearest-neighbour correlation along $y$, while along $x$ we find antibunching caused by the Pauli repulsion of the holes (see Fig.\,\ref{fig:fig2}a). Furthermore, at larger distances $d_y>1$ there are positive correlations, which indicates that, instead of merely forming isolated, nearest-neighbour hole pairs, there is a finite probability that vertically aligned hole structures are extended through the system.  Additionally, there are significant correlations along the diagonals at $\dvec=(1,1)$, which we interpret as signs of charge fluctuations along $x$ due to the finite coupling $t_x$. The correlations at $d_y=2$ are slightly suppressed which we attribute to next-nearest-neighbour hopping (see \hyperref[sec:mixD]{SI}). Finally, the positive signal at $d_x=\pm5$ may be related to the presence of a second, vertically aligned charge structure in the system reminiscent of a charge density wave.

By considering 1d cuts along $y$ and $x$ (Fig.\,\ref{fig:fig2}b) we corroborate the bunching (antibunching) along $y$ ($x$) through the system in the mixD setting. For the standard 2d system ($\Delta=0$, Fig.\,\ref{fig:fig2}b inset), in contrast, there is antibunching along both directions. The anticorrelations along $x$ are enhanced by removing $t_y$ due to the absent competition between anticorrelations along $x$ and $y$. 

In order to identify whether an ideal doping level for the emergence of stripes exists in our mixD system, we bin our data by doping and calculate \gh per bin (see Fig.\,\ref{fig:fig2}c). Both the nearest-neighbour and diagonal correlations decrease with doping, indicative of the decrease in pairing probability with doping and compatible with a reduction of the spin correlations responsible for the binding. For $\dvec=(0,2), (0,3)$ there is a non-trivial dependence on doping with positive correlator values starting at $\delta=0.17$. This is indicative of a possible transition from the formation of individual pairs to extended stripes~\cite{Huang2018,Zheng2017,Wietek2021}.

We compare the correlations along $y$ to DMRG calculations of Eq.\,\eqref{eq:tJ} on $8\times3$ sites, $J_y/t_x = 0.5$, $\kB T/t_x = 0.41$ as a function of doping in Fig.\,\ref{fig:fig2}d (see also \hyperref[sec:numerics]{SI} for normalisation). While qualitatively the result is comparable to the experimental data, quantitative differences are expected due to the strong finite size limitations in the DMRG along $y$. Further differences could arise due to the presence of the aforementioned second-order hopping process which introduces additional repulsion between holes as well as the statistical distribution of holes between different chains in the experiment while calculations feature balanced hole numbers. 

\section{Structures beyond hole pairing}
\begin{figure}[!t]
	\centering
	\includegraphics[scale = 1]{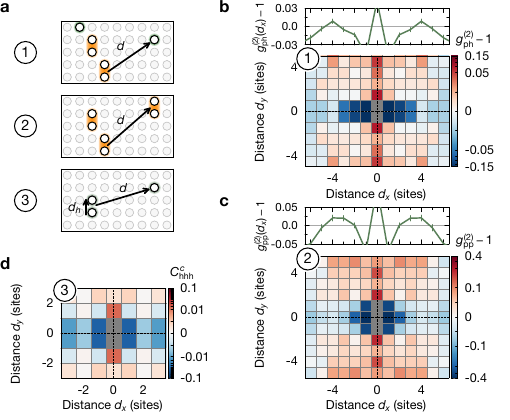}
	\caption{\textbf{Multi-point correlators.} \textbf{a}, Illustration of pair-hole, pair-pair and hole-hole-hole correlator, where a pair is defined as a nearest-neighbour pair of holes along $y$. \textbf{b} (\textbf{c}), Symmetrised correlation map of the pair-hole (pair-pair) correlator. We find an attraction of the pairs along $y$ which points towards the formation of larger-scale structures. Above the map, the average over $d_y$ hints at the existence of another charge structure at $d_x=4$. In the symmetrised, connected three-point hole-hole-hole correlator with $\dvec^h=(0,1)$ we observe a positive signal at nearest neighbours along $y$ in \textbf{d} which indicates the existence of longer charge structures beyond pairs of two holes (see \hyperref[sec:map_errs]{SI} for statistical significance). The data is evaluated over the hole doping distribution as given in the \hyperref[sec:hist]{SI}.}
	\label{fig:fig3}
\end{figure}
The connected two-point correlator \gh only provides limited insights into the physics of extended charge structures and how they interact with each other. We extend the analysis by considering the two-point pair-hole and pair-pair correlators 
\begin{flalign}\label{eq:ph}
	\begin{aligned}
	&g^{(2)}_{\mathrm{ph}}(\dvec) - 1 = \frac{1}{\mathcal{N_\dvec}}\sum_{\ivec} \left(\frac{\langle \hat{n}^p_\ivec \hat{n}^h_{\ivec+\dvec} \rangle}{\langle \hat{n}^p_\ivec\rangle \langle \hat{n}^h_{\ivec+\dvec} \rangle} - 1\right),\\
	&g^{(2)}_{\mathrm{pp}}(\dvec) - 1 = \frac{1}{\mathcal{N_\dvec}}\sum_{\ivec} \left(\frac{\langle \hat{n}^p_\ivec \hat{n}^p_{\ivec+\dvec} \rangle}{\langle \hat{n}^p_\ivec\rangle \langle \hat{n}^p_{\ivec+\dvec} \rangle} - 1\right),
	\end{aligned}
\end{flalign}
where we define the pair operator $\hat{n}^p_{\ivec}=\hat{n}^h_{(i_x,i_y)}\hat{n}^h_{(i_x, i_y+1)}$. These correlators, as illustrated in Fig.\,\ref{fig:fig3}a, assume the existence of nearest-neighbour pairs along $y$ (established using \gh) and consider the attraction or repulsion of these pairs to other dopants or pairs. They may be seen as fully connected and normalised two-point correlators of pairs or `partially connected' three/four-point hole correlators. Note that for simplicity we neglect diagonal pairs (i.e. $\hat{n}^h_{(i_x,i_y)}\hat{n}^h_{(i_x\pm1, i_y+1)}$), associated with fluctuations along $x$, and may thus underestimate the amount of order within the system.

We present the pair-hole and pair-pair correlations for the mixD system as a function of distance along $x$ and $y$ in Fig.\,\ref{fig:fig3}b and c. For improved statistics, we include in our analysis all hole doping levels (see \hyperref[sec:hist]{SI}) for which the offset $o_\delta$ of Eq.\,\eqref{eq:hh} becomes negligible. In both cases, we observe positive correlations along $y$ which extend throughout the system, indicating that individual pairs are not repelled from other holes or each other but instead align along $y$ and tend to form stripe-like structures. Meanwhile, there is a strong anticorrelation along $x$ for $|d_y|\leq1$, which we attribute to the antibunching of individual holes in the same chain. We also compute the average of the correlators over $d_y$ (top of Fig.\,\ref{fig:fig3}b, c). This reveals a slightly positive signal at a distance of $d_x=4$ which, similar to Fig.\,\ref{fig:fig2}, may be related to the onset of a charge density wave.

For further insights into the binding of larger structures, we consider the fully connected three-point hole-hole-hole correlator
\begin{equation}\label{eq:hhh}
	C^{c}_\mathrm{hhh}(\dvec^h,\dvec) = \frac{1}{\mathcal{N}_{\dvec^h,\dvec}}\sum_{\mathclap{\substack{\ivec\\\jvec = \ivec+\dvec^h/2+\dvec}}} \left(\frac{\langle \hat{n}^h_\ivec \hat{n}^h_{\ivec+\dvec^h} \hat{n}^h_{\jvec}\rangle - C_\mathrm{disc}}{\langle \hat{n}^h_\ivec\rangle \langle \hat{n}^h_{\ivec+\dvec^h}\rangle\langle \hat{n}^h_{\jvec} \rangle}\right)
\end{equation}
where $C_\mathrm{disc}$ removes all lower-order disconnected parts of the correlator (see \hyperref[sec:corrs]{SI}). We show the correlator for $\dvec^h=(0,1)$ in Fig.\,\ref{fig:fig3}d and find a positive signal at the closest distance along $y$ while all other distances are negative (along $x$) or vanish within the error bars (see \hyperref[sec:map_errs]{SI}). This signal directly points to extended charge structures being favoured in excess of just individual hole pairs.\\

\begin{figure}[t]
	\centering
	\includegraphics[scale = 1]{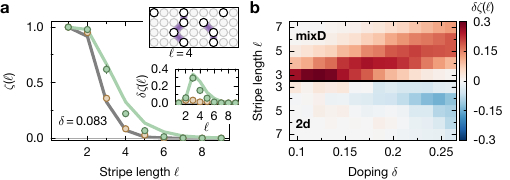}
	\caption{\textbf{Stripe length histograms.} \textbf{a} Normalised histogram of `stripes' (as defined in the text) of at least length $\ell$ in mixD (green) and 2d systems (brown) at a doping of $\delta = 0.083$ compared to a random distribution of holes (grey line). We compare to a mean field theory (see \hyperref[sec:MFT]{SI}) at $\kB T/t_x=0.355$ (light green line). The inset shows the difference $\delta\zeta(\ell)$ to the random distribution (see \hyperref[sec:random]{SI}). The full doping dependence of $\delta\zeta(\ell)$ is shown in \textbf{b} where the excess occurrences tend towards longer lengths with doping.}
	\label{fig:fig5}
\end{figure}

In order to provide additional evidence for extended, fluctuating charge structures, we make use of the full information in our snapshots and count `stripes'. To this end, we define a fully filled `stripe' as a connected line of holes along $y$, where the pairwise distance along $x$ between holes in neighbouring chains is at most 1 (see Fig.\,\ref{fig:fig5}a inset).
We designate the length $\ell$ of this structure by the number of chains involved. We then consider the fraction $\zeta(\ell)$ of experimental realisations containing a `stripe' of at least length $\ell$.
In Fig.\,\ref{fig:fig5}a we compare the mixD (green) case with the 2d system (brown) and randomly distributed holes (grey line, see \hyperref[sec:random]{SI}) at a doping of $\delta = 0.083$ analysed on a subsystem of $9\times9$ sites. For mixD we find an excess of events for large $\ell$ consistent with the tendency to form long fluctuating structures while the results obtained for the standard 2d case are consistent with randomly distributed holes. Full numerical calculations are out of reach at our system size and temperature range, but a mean-field model of stripes shows quantitative agreement in the low doping regime (see green lines in Fig.\,\ref{fig:fig5}a and \hyperref[sec:numerics]{SI}). We next analyse the difference to the random distribution $\delta\zeta(\ell)$ as a function of doping (Fig.\,\ref{fig:fig5}b). For all doping levels and lengths, this signal is positive in the mixD system, indicating the inclination of the system to form extended structures. The excess probability at longer lengths grows with doping as structures of increasing lengths form.

\section{Spin sector}
\begin{figure}[t]
	\centering
	\includegraphics[scale = 1]{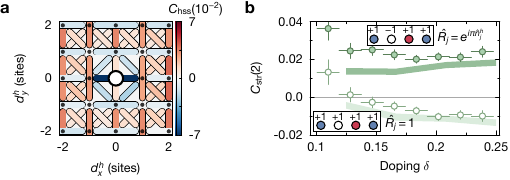}
	\caption{\textbf{Spin sector analysis.} \textbf{a}, Hole-spin-spin correlation map. We show the bare correlator for diagonal and selected next-nearest spin-bonds as a function of distance from the hole. The strongest signal is found in the sign change of the next-nearest neighbour bond across the hole along $x$ pointing towards a domain wall in the local AFM pattern. Along $y$, the correlations keep their expected positive sign from the AFM pattern. \textbf{b}, Similarly, by considering the string spin correlator (dark green) and normal spin-spin correlator (light green) at distance $d_x=2$ (see text), we observe a change in sign consistent with the change in parity of the AFM pattern. Shaded regions are theory results at $\kB T/t_x=0.3$. Error bars are estimated using bootstrapping and are smaller than the marker size if not visible.}
	\label{fig:fig4}
\end{figure}
The AFM correlations in the system and their interplay with charge delocalisation is crucial for the formation of stripes and leads to characteristic signatures in the spin sector\cite{Schulz1989}. Most prominently, one expects a change in the parity of the AFM order in the presence of stripes, manifesting as incommensurate magnetism of the system and splitting of the peak at $(\pi,\pi)$ in the spin structure factor \cite{Kivelson2003} (as also known in 1d systems~\cite{Salomon2019}). While our parameter regime is not favourable to investigate structure factors (see \hyperref[sec:sequence]{SI}), our microscopic resolution in both spin and charge sector allows us to evaluate real-space observables inaccessible in solid-state experiments. Most useful in this context are higher-order spin-charge correlators such as the normalised, bare 3-point hole-spin-spin correlator
\begin{equation}\label{eq:eq4}
	C_\mathrm{hss}(\dvec^s,\dvec^h) = \frac{1}{\mathcal{N}_{\dvec^s,\dvec^h}} \sum_{\mathclap{\substack{\ivec\\\jvec = \ivec+\dvec^h-\dvec^s/2}}} \frac{\langle \hat{n}^h_\ivec \hat{S}^z_{\jvec} \hat{S}^z_{\jvec+\dvec^s}\rangle}{\langle \hat{n}^h_\ivec \rangle \sigma(\hat{S}^z_{\jvec}) \sigma(\hat{S}^z_{\jvec+\dvec^s})},
\end{equation}
where $\dvec^s$ is the spin bond vector, $\dvec^h$ the distance of the bond from the dopant and we normalise by hole density $\langle \hat{n}^h\rangle$ and the spin standard deviation $\sigma(\hat{S}^z)$.

Previous studies have shown that, in square lattice 2d Fermi-Hubbard systems, a single mobile dopant will be surrounded by a dressing cloud of reduced spin correlations forming a magnetic polaron~\cite{Koepsell2019, Hartke2020, Koepsell2021}.  In 1d systems, incommensurate magnetism leads to a change in the parity of the AFM pattern across impurities~\cite{Hilker2017}. The same feature is predicted to prevail in stripe phases, making this correlator suited to revealing this specific feature in our data. We show the bare hole-spin-spin correlator as defined in Eq.\,\eqref{eq:eq4} for the mixD system in Fig.\,\ref{fig:fig4}a, where specific spin bonds are shown for varying distances from a hole. We focus on the diagonal and next-nearest neighbour correlators. The most prominent feature is the strongly negative correlation across the hole along $x$, which is consistent with a change in the parity of the local AFM pattern across a hole. Similarly, the diagonal bonds in the direct vicinity of the hole also become negative. This is another indication of fluctuations along $x$ within charge structures. Meanwhile, the $\dvec^s=(0,2)$ correlations along $y$ are largely unaffected by the presence of a hole and retain their positive sign. The slightly negative (positive) $\dvec^s=(2,0)$ ($\dvec^s=(1,1)$)-bond in the background further away from the dopant is due to the overall doping level and vanishes in the connected correlator (see \hyperref[fig:3point]{SI}).\\

Another way to elucidate the change in spin order across dopants employs a spin-string correlator~\cite{Kruis2004,Hilker2017}. This spin-spin correlator has additional sign changes for every hole between two spins in the same chain and is defined as
\begin{equation}\label{eq:squeezed}
	C_\mathrm{str}(d) =\frac{1}{\mathcal{N}_d}\sum_i \frac{\left< \hat{S}^z_i\left(\prod_{j=1}^{d-1}\hat{R}_{i+j}\right)\hat{S}^z_{i+d}\right> - \left< \hat{S}^z_i \right> \left< \hat{S}^z_{i+d} \right> }{\sigma(\hat{S}^z_i)\sigma(\hat{S}^z_{i+d})},
\end{equation}
where $\hat{R}_i = e^{i\pi \hat{n}^h_i}$. Note that for $\hat{R}_i=\mathds{1}$ the common spin-spin correlator is recovered. For systems with spin-charge separation, this correlator reveals a hidden spin structure in doped AFM systems~\cite{Hilker2017}. The changes in the phase of the AFM pattern for stripe phases act in a similar fashion and can be revealed by measuring this string correlator along the direction perpendicular to the stripes (i.e. along $x$). We show both correlators at distance $d=2$ in Fig.\,\ref{fig:fig4}b as a function of doping. We observe a change to a positive sign upon employing the string correlator that only varies weakly with doping, in agreement with theory predictions. These features can be directly related to the characteristic spin domain parity flips present in stripe phases. Note that we observe these features even without long-range AFM correlations - which are only expected at lower temperatures - because stripe-like structures already energetically favour such a local spin arrangement.

\section{Conclusion}
We have realised a mixed-dimensional Fermi-Hubbard model using ultracold atoms and found signatures of hole pairing and extended charge ordering in a temperature regime with short-ranged spin correlations, where the collective behaviour of charges remains poorly understood. We detect effective hole attraction in density correlations and present additional evidence for the onset of fluctuating stripes and their interplay with the magnetic background using real-space observables. 
In addition, the spin environment around such stripe-like structures is in qualitative agreement with the formation of an AFM domain wall across the dopants both in 3-point and string correlators. We interpret these features as signatures of stripe formation in a cold-atom Fermi-Hubbard system. As lower temperatures become available in experiments, the same analysis shown here could be carried out, paving the way towards more detailed studies of stripes, extracting their precise periodicity, fluctuations and filling. Thanks to the favourable energy scales of the mixD setting, quantum simulations are now in a position to study collective phenomena and provide valuable comparisons to recent results in theoretical calculations~\cite{Qin2020,Xiao2023}. 
Via the mapping to attractive interactions~\cite{Ho2009}, new insights into the stripe phase also directly relate to the exotic FFLO phase~\cite{Moreo2007}. 
Direct extensions to bilayer mixed-dimensional systems furthermore connect our work to recently discovered high-$T_c$ compounds, where the mixed dimensionality seems essential for the emergence of a superconducting phase at around $80\,\unit{\kelvin}$ in bilayer nickelates~\cite{Qu2023,Sun2023,Schloemer2023}.

\textbf{Acknowledgments}:
We thank E. Demler for fruitful discussions. This work was supported by the Max Planck Society (MPG), the Horizon Europe programme HORIZON-CL4-2022 QUANTUM-02-SGA (project 101113690, PASQuanS2.1), the German Federal Ministry of Education and Research (BMBF grant agreement 13N15890, FermiQP) and Germany’s Excellence Strategy (EXC-2111-390814868). T.C. acknowledges funding from the Alexander v. Humboldt Foundation. F.G. acknowledges funding from the European Research Council (ERC) under the European Union’s Horizon 2020 research and innovation programme (grant agreement no. 948141) from ERC Starting Grant SimUcQuam.

\textbf{Author contributions}: D.B. led the project. T.C. and D.B. contributed significantly to data collection and analysis. H.S., A.B. and F.G. performed the theory calculations. D.B. and T.A.H. wrote the manuscript. T.A.H. and I.B. supervised the study. All authors worked on the interpretation of the data and contributed to the final manuscript.

\textbf{Competing interests}: The authors declare no competing interests.\\

\bigskip

\newpage

\clearpage
\newpage
\makeatletter
\renewcommand{\thefigure}{S\@arabic\c@figure}
\newcommand{\bb}[1]{{\mathbf #1}}
\newcommand{\bs}[1]{{\boldsymbol #1}}

\makeatother
\setcounter{figure}{0}
\setcounter{table}{0}
\setcounter{section}{0}

\section*{Supplementary Information}

\subsection{Experimental sequence}\label{sec:sequence}
\begin{figure}[b]
	\centering
	\includegraphics[scale = 1]{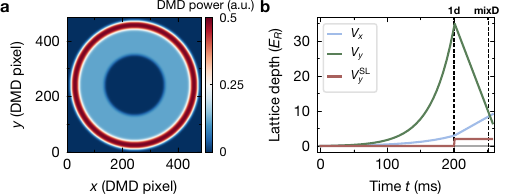}
	\caption{\textbf{Lattice potential and ramps.} \textbf{a} Pattern applied to the DMD for potential shaping. \textbf{b}, Lattice ramps to prepare the mixD system. We first ramp in $200\,\unit{\milli\second}$ to decoupled 1d chains before ramping to the full mixD system.}
	\label{fig:ramps}
\end{figure}

We prepare a spin-balanced sample of ultracold $^6$Li atoms in the two lowest hyperfine states $|F=1/2,m_F=\pm1/2\rangle$ in a single layer of an optical lattice following our previous work~\cite{Hirthe2023,Koepsell2020}. After magnetic evaporation, we load from a crossed dipole trap into a box potential (surrounded by a reservoir with $\sim h\times2 \,\unit{\kilo\hertz}$ higher chemical potential) projected with a digital mirror device (DMD, see Fig.\,\ref{fig:ramps}a). From this, we load into optical lattices along $x$ and $y$ with $a_x=1.11\,\unit{\micro\meter}$, $a_y=1.14\,\unit{\micro\meter}$. Their depths in the following are given in units of their respective lattice recoil $E_R = h^2/(8Ma^2)$ where $M$ is the atomic mass.

In order to prepare the mixed-dimensional system described in the main text without introducing large density inhomogeneities, we cannot directly load into the final lattice configuration but instead follow a procedure similar to~\cite{Hirthe2023} (see Fig.\,\ref{fig:ramps}). We first load into decoupled 1d chains along $x$ by exponentially ramping to $V_x = 3\,E_R$, $V_y=35\,E_R$ and a scattering length of $1160\,a_\mathrm{B}$, corresponding to our final on-site interactions of $U=h\times4.4(1)\,\mathrm{kHz}$, within $200\,\mathrm{ms}$. At this point we turn on a superlattice along $y$ ($a_y^{\mathrm{SL}}=2a_y=2.28\,\unit{\micro\meter}$) to a depth of $V_{y}^{\mathrm{SL}} = 2\,E_R$ within $1\,\unit{\milli\second}$. By tuning the relative phase between the lattice and the superlattice to the fully staggered configuration, we ensure that the spin couplings remain the same in even and odd bonds along $y$. This staggering creates a potential offset of $\Delta=0.65(5)\,U$ between neighbouring sites to suppress tunnelling along $y$. We then slowly restore coupling along $y$ by ramping the lattices in $56\,\unit{\milli\second}$ to their final depths of $V_x=9\,E_R$, $V_y=7\,E_R$. We make sure to keep the interactions constant during this second ramp by adjusting the scattering length accordingly, leading to a final scattering length of $1293\,a_\mathrm{B}$. For any 2d system comparison, we perform the same ramps without turning on the $y$-superlattice. For more details on our superlattice design see~\cite{Chalopin2023}. For detection we freeze out the system by ramping to $V_{x/y}=43.5E_R$ within $1.5\,\unit{\milli\second}$ and perform spin-resolved single-site detection as described in~\cite{Hirthe2023}.

The resulting system can be accurately described by a Fermi-Hubbard-type model with parameters ($t$, $U$, $\Delta$) which can be mapped onto the $t$-$J$-model of Eq.\,\eqref{eq:tJ}. For all settings we have tunnelling $t_x=h\times163(10)\,\mathrm{\unit{\hertz}}$, interactions $U=h\times4.4(1)\,\mathrm{\unit{\kilo\hertz}}$ (thus $U/t_x=27(2)$) and superexchange $J_x=h\times24(4)\,\mathrm{\unit{\hertz}}$. For the 2d system we have $t_y^{\mathrm{2d}}=h\times253(13)\,\mathrm{\unit{\hertz}}$, however for $\Delta\neq 0\,U$ the effective coupling $t_y$ is negligible. The superexchange coupling $J_y$ is nonzero in both cases with $J_y^{2d}=h\times58(7)\,\mathrm{\unit{\hertz}}$ for the 2d system and $J_y=h\times104(23)\,\mathrm{\unit{\hertz}}$ for $\Delta=0.65(5)\,U$.
Due to the strongly anisotropic spin couplings and large $U/t_x$, the spin correlations are not sufficiently long-ranged to expect any signal in the spin structure factor.

\begin{figure}[t]
	\centering
	\includegraphics[scale = 1]{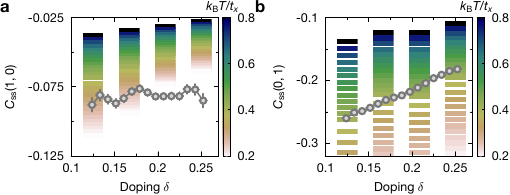}
	\caption{\textbf{Spin correlations as a function of doping.} Nearest neighbour spin correlations along $x$ (\textbf{a}) and $y$ (\textbf{b}) for different doping levels. We compare the experimental data (grey markers) to numerical data for $C_\mathrm{ss}(1,0)$ for different temperatures for simulations on $L_x,L_y = 8,3$ and $J_y/t_x = 0.5$ to get an estimate for our temperature.}
	\label{fig:ss_doping}
\end{figure}

We estimate the temperature of our system using the spin correlations $	C_\mathrm{ss}(\dvec) =\frac{1}{\mathcal{N}_\dvec}\sum_i \frac{\left< \hat{S}^z_\ivec \hat{S}^z_{\ivec+\dvec}\right>  - \left< \hat{S}^z_\ivec \right> \left< \hat{S}^z_{\ivec+\dvec} \right> }{\sigma(\hat{S}^z_\ivec)\sigma(\hat{S}^z_{\ivec+\dvec})}$ as a function of doping and compare this to MPS calculations in Fig.\,\ref{fig:ss_doping}. We fit the individual doping bins to the numerical data and extract their respective temperature. As the short $y$-direction may be subject to finite size effects in the DMRG calculations, we determine individual temperatures along $x$ and $y$. By extracting temperatures per doping level, we estimate a temperature of $\kB T/t_x \approx 0.3(1)$ and $\kB T/t_x \approx 0.4(1)$ from the correlations along $x$ and $y$ respectively.

\subsection{Offset phase calibration}\label{sec:phasescan}
\begin{figure}[t]
	\centering
	\includegraphics[scale = 1]{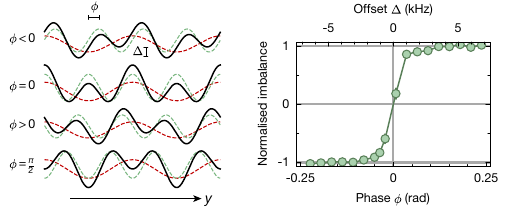}
	\caption{\textbf{Phase calibration.} We load a dilute cloud into a deep superlattice ($V_x=40\,E_R$, $V_y=8\,E_R$ and $V_{y}^{\mathrm{SL}} =21\,E_R$) with different phase and extract the imbalance in occupation between neighbouring chains. At the symmetric double well configuration ($\phi=0$) we reach zero imbalance while even for small deviations we quickly occupy only one part of the double well. All main experimental results are obtained for $\phi=\pi/2$.}
	\label{fig:phasescan}
\end{figure}
In order to calibrate the detuning $\Delta$, we first need to precisely determine the relative phase between the lattice and the superlattice. For this, we load a dilute cloud into a system of decoupled double wells along $y$ where $V_x=40\,E_R$, $V_y=8\,E_R$ and $V_{y}^{\mathrm{SL}} =21\,E_R$ leading to an intrawell coupling of $t_y(\phi=0) = h\times724(80)\,\unit{\hertz}$. We vary the phase between the lattice and the superlattice and measure the normalised imbalance, i.e. the difference in occupation between the different parts of the double well, normalised by their summed occupation (see Fig.\,\ref{fig:phasescan}). When we prepare symmetric double wells, the imbalance approaches zero. However, when we tune away from this configuration, we reach an imbalance of $\pm1$ within less than $50\,\unit{\milli\radian}$. This sharp transition indicates a high degree of stability and homogeneity of the relative superlattice phase within the system (see also~\cite{Chalopin2023} for more details). For the measurements presented here, we then work at a phase of $\phi=\pi/2$, where the offset between neighbouring lattice sites is highest for a given lattice depth and the interwell and intrawell couplings are identical. 

\begin{figure}[t]
	\centering
	\includegraphics[scale = 1]{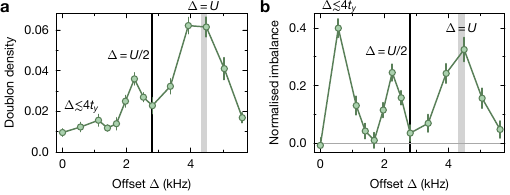}
	\caption{\textbf{Offset scan.} Doublon density (\textbf{a}) and imbalance between chains (\textbf{b}) as a function of potential offset $\Delta$ (i.e. superlattice power) for a relative superlattice phase of $\pi/2$. The peak in the doublon density coincides with the interaction energy $U$ (grey line) at which atoms are then resonantly transferred to neighbouring chains. For small offsets, tunnelling is not yet fully suppressed and an imbalance is created. Above an intermediate peak at $U/2$ (created by a higher order process), there is a low imbalance regime where the experiment is performed (black line).}
	\label{fig:tiltscan}
\end{figure}

We confirm the energy scales associated with a given potential offset by comparing it to our interaction energy. We prepare the system at the lattice parameters stated in the previous section and a phase of $\phi=\pi/2$ and vary the depth of the superlattice (i.e. the potential offset) in a slightly hole doped system (see Fig.\,\ref{fig:tiltscan}). For offsets smaller than the band width, tunnelling between sites is not yet suppressed and we create a strong imbalance. On the other hand, for large offsets around the interaction energy, we enable resonant tunnelling between sites whenever both are occupied, thus creating both an imbalance and doublons within the system. The observed scales are consistent with band structure calculations based on our lattice parameters. Between these two regimes, the imbalance approaches zero. The maximum in imbalance around $U/2$ can be explained by a second-order process where a doublon in a lower chain breaks into two atoms in the two adjacent chains (see also~\cite{Chalopin2023}). To avoid this effect and the associated additional holes and doublons, we perform our experiments at an offset slightly above $U/2$.

\subsection{Data statistics, doping histograms}\label{sec:hist}
\begin{figure}[b]
	\centering
	\includegraphics[scale = 1]{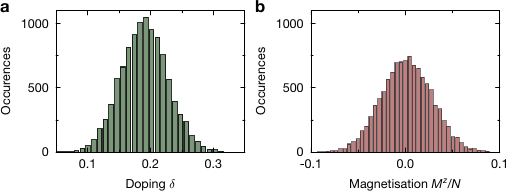}
	\caption{\textbf{Data statistics.} Histograms of doping (\textbf{a}) and magnetisation (\textbf{b}). We take data between 10 and 30\% doping while the total magnetisation is well centred around 0.}
	\label{fig:stats}
\end{figure}

In total, we collect 11675 experimental realisations. Out of these, 1254 were taken in a 2d system with $\Delta=0\,U$, the remaining 10421 with $\Delta=0.65(5)\,U$. Within these measurements, we slightly vary the doping level (additionally to the natural fluctuations inherent to our preparation scheme) which yields a range of $10-30\%$ hole doping. To ensure that there is no overall magnetisation $M^z = \sum_i \langle S_i^z\rangle$ within the system, we check the distribution of magnetisation normalised by the system size, which is centred around zero and shows a width below shot-noise (see Fig.\,\ref{fig:stats}).

\subsection{Connected correlators and offsets}\label{sec:corrs}
\subsubsection{Full connected correlator expressions}
\begin{figure}[t!]
	\centering
	\includegraphics[scale = 1]{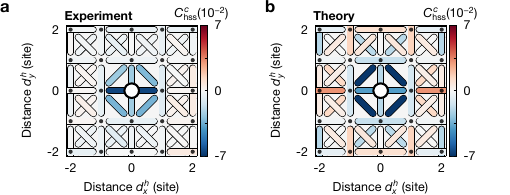}
	\caption{\textbf{Connected 3-point correlator.} \textbf{a}, Fully connected, symmetrised, three-point hole-spin-spin correlator. By removing the AFM background we focus on the additional effect introduced by the dopant which is compatible with the onset of a domain wall in the local AFM pattern across the dopant. A comparison to DMRG calculations at $\kB T/t_x=0.4$ (\textbf{b}) shows qualitatively similar results.}
	\label{fig:3point}
\end{figure}
We present a variety of correlators to characterise the spin and charge order in our system. We here distinguish between bare, `partially connected' and fully connected correlators. 
While the bare correlator does not subtract anything, the fully connected correlator subtracts all possible lower order contributions between all its constituents, e.g. for a two-point correlator it removes the product of the mean operator values while for a three-point correlator it also removes all combinations of two-point correlators. Meanwhile, `partially connected' correlators only subtract some specific lower-order correlators: in this case, as we consider pairs as new objects, we don't subtract any correlations stemming from the individual holes in the pair. 

All these different types of correlators are then helpful to extract slightly different information about the system. While fully connected correlators are especially useful to extract small signals in higher-order correlators dominated by lower-order contributions, the bare correlator may be more interesting when higher-order correlations are actually larger than lower-order correlators.

For this reason, in addition to the `partially connected' pair-hole and pair-pair correlator (Eq.\,\eqref{eq:ph}) and the bare hole-spin-spin correlator (Eq.\,\eqref{eq:eq4}), we used the fully connected hole-hole-hole correlator defined as
\begin{multline}\label{eq:hhh_full}
	C^c_\mathrm{hhh}(\dvec^h,\dvec) = \frac{1}{\mathcal{N}_{\dvec^h,\dvec}} \sum_{\substack{\ivec\\ \jvec=\ivec + \dvec^h/2 +\dvec}} \frac{1}{\langle \hat{n}^h_\ivec \rangle \langle \hat{n}^h_{\ivec+\dvec^h} \rangle \langle\hat{n}^h_\jvec\rangle}\times\\
	\left( \langle \hat{n}^h_\ivec \hat{n}^h_{\ivec+\dvec^h} \hat{n}^h_\jvec\rangle 
	- \langle \hat{n}^h_\ivec \hat{n}^h_{\ivec+\dvec^h}\rangle \langle \hat{n}^h_\jvec \rangle 
	- \langle \hat{n}^h_\ivec \rangle \langle\hat{n}^h_{\ivec+\dvec^h} \hat{n}^h_\jvec \rangle \right. \\
	\left. - \langle \hat{n}^h_\ivec \hat{n}^h_\jvec\rangle \langle \hat{n}^h_{\ivec+\dvec^h} \rangle 
	+ 2\langle \hat{n}^h_\ivec\rangle\langle \hat{n}^h_{\ivec+\dvec^h}\rangle\langle \hat{n}^h_\jvec\rangle\right).
\end{multline}
Similarly, we can define a connected hole-spin-spin correlator as 
\begin{multline}\label{eq:hss_conn}
	C^c_\mathrm{hss}(\dvec^s,\dvec^h) = \frac{1}{\mathcal{N}_{\dvec^s,\dvec^h}} \sum_{\mathclap{\substack{\ivec \\ \kvec-\jvec = \dvec^s,\\ (\kvec+\jvec)/2-\ivec = \dvec^h}}}
	\frac{1}{\langle \hat{n}^h_\ivec \rangle \sigma(\hat{S}^z_\jvec) \sigma(\hat{S}^z_\kvec)}\times\\
	\left(\langle \hat{n}^h_\ivec \hat{S}^z_\jvec \hat{S}^z_\kvec\rangle\right.\\
	- \langle \hat{n}^h_\ivec \rangle\langle\hat{S}^z_\jvec \hat{S}^z_\kvec\rangle
	- \langle \hat{n}^h_\ivec \hat{S}^z_\jvec \rangle\langle\hat{S}^z_\kvec\rangle
	- \langle \hat{n}^h_\ivec \hat{S}^z_\kvec \rangle\langle\hat{S}^z_\jvec\rangle \\
	\left.+ 2\langle \hat{n}^h_\ivec\rangle\langle \hat{S}^z_\jvec \rangle\langle\hat{S}^z_\kvec\rangle\right).
\end{multline}
For this connected correlator, we observe the same main features also shown in Fig.\,\ref{fig:fig4}a in the main text with a dominant negative bond across the hole (see Fig.\,\ref{fig:3point}). This signal is strong enough to dominate over the AFM background, changing the correlator sign even in the bare correlator shown in Fig.\,\ref{fig:fig4}. Meanwhile, the positive diagonal and next-nearest neighbour bonds along $y$ far from the hole shown in Fig.\,\ref{fig:fig4}a now vanish as they are not related to the presence of a hole but just stem from the AFM background. We compare this to DMRG calculations with $L_y=4$, $\delta=0.125$, $\kB T/t_x=0.4$ which shows the same main features of strong anticorrelations across the dopant and at the diagonals in the immediate vicinity.
\subsubsection{Offset correction}
In addition to the subtraction of the disconnected part, we also introduce an offset correction $o_\delta$ on the hole-hole correlator. This correction arises due to the doping fluctuations in our finite-sized system. For each realisation, we prepare a system with random but fixed total atom number and magnetisation (see Fig.\,\ref{fig:stats}). The calculated correlations in a finite system then obey a sum rule depending on the particle number and variance.

We start by considering $N$ fermions on $V$ sites with density $n=N/V$. The local two-point correlator $\Gamma(i,j)=\frac{\langle \hat{n}_i \hat{n}_j\rangle}{n_i n_j} -1$ (with $n_i= \langle \hat{n}_i\rangle$) after summing over all possible pairs of sites $i,j$ can be expressed as
\begin{equation}
	\sum_{i,j} \Gamma(i,j) = \sum_{i,j} \left(\frac{\langle \hat{n}_i \hat{n}_{j}\rangle}{n_i n_{j}}-1\right) \approx \left( \frac{\langle \hat{N}^2\rangle}{n^2}-V^2\right),
\end{equation}
where we used  $\hat{N}^2=\sum_{i,j} \hat{n}_i\hat{n}_j$ and $n_i\approx n_{j}\approx n$ in our homogeneous system. This we can relate to the variance as
\begin{equation}
	\frac{1}{V^2}\sum_{i,j} \Gamma(i,j) = \frac{\mathrm{Var}(\hat{N})}{N^2}.
\end{equation}
\begin{figure}[tb!]
	\centering
	\includegraphics[scale = 1]{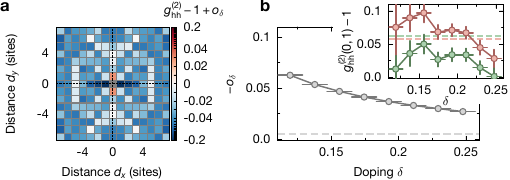}
	\caption{\textbf{Correlator offsets.} \textbf{a}, Correlation map of Fig.\,\ref{fig:fig2}a without offset correction. \textbf{b}, Correlator offset $o_\delta$ as a function of doping. The nearest neighbour hole-hole correlator as a function of doping with (red) and without (green) offset correction is shown in the inset. The horizontal dashed lines are the same correlator without binning by density in which case the offset almost vanishes (dashed line in \textbf{b}).}
	\label{fig:offset}
\end{figure}
If we now separate the on-site fluctuations and use fermionic statistics where $\hat{n}^2=\hat{n}$ (and thus $\Gamma(i,i) =\frac{1}{n} - 1$), we obtain 
\begin{equation}\label{eq:sumrule}
	\frac{1}{V^2}\sum_{i\neq j} \Gamma(i,j) = \frac{\mathrm{Var}(\hat{N})-N(1-n)}{N^2}. 
\end{equation}
Unless the global fluctuations of $N$ are also fermionic fluctuations (i.e. multinomial where $\mathrm{Var}(\hat{N}) = N(1-n)$), the sum rule in Eq.\,\eqref{eq:sumrule} leads to a nonzero value of $\Gamma(i,j)$ for $i\neq j$ even at $T=\infty$. Note that typically $\mathrm{Var}(\hat{N}) \sim N$ or less, such that Eq.~\eqref{eq:sumrule} is a $1/N$ correction, which vanishes in the thermodynamic limit.

Identifying $\hat{n}\equiv\hat{n}^h$, we use this result in the calculation of the hole-hole correlations in Fig.\,\ref{fig:fig2} and thus define the offset $o_\delta$ via
\begin{equation}\label{eq:offset}
	o_\delta = \frac{\mathrm{Var}(\hat{N}^h)-N^h(1-n^h)}{(N^h)^2},
\end{equation}
and the corrected correlation as Eq.~\eqref{eq:hh} of the main text
\begin{equation}
	g^{(2)}_{\mathrm{hh}}(\dvec) -1 = \frac{1}{\mathcal{N_\dvec}}\sum_{\ivec} \left(\frac{\langle \hat{n}^h_\ivec \hat{n}^h_{\ivec+\dvec} \rangle}{\langle \hat{n}^h_\ivec\rangle \langle \hat{n}^h_{\ivec+\dvec} \rangle} - 1 - o_\delta\right),
\end{equation}
with $\mathcal{N}_\dvec = \sum_{\ivec,\jvec} \delta_{\jvec,\ivec+\dvec}$ and Kronecker delta $\delta_{\ivec,\jvec}$.
Most importantly, the doping dependent offset we apply is global on all distances.

This offset correction $o_\delta$ only plays a role when selecting specific doping levels in a finite size system such that the total atom number is almost fixed ($\mathrm{Var}(\hat{N})\to 0$) and thereby leads to strong global offsets, that we hereby compensate (see Fig.\,\ref{fig:offset}a). We show in Fig.\,\ref{fig:offset}b the offset as a function of doping together with the nearest neighbour hole-hole correlator values with and without applied offset. As indicated by the dashed lines, the offset without selection on a density bin is negligible. For this reason, we do not apply any corrections in Fig.\,\ref{fig:fig3}.

\subsubsection{Correlator from theory}
When comparing the absolute values of hole-hole correlations to simulations, care needs to be taken due to the differences in doping, fluctuations and boundary conditions. All calculations are performed with open boundary conditions along $x$ and $y$. Meanwhile, the potential at the edges in the experiment has a finite width which means that the exact position of any charge feature will be fluctuating and therefore be washed out. As a result, we detect signals in \gh but not in the density, in contrast to theory where stripes show up as density features~\cite{Hirthe2023}. When using connected correlators on theory data, this will lead to reduced correlations. To analyse numerical results, we hence use the slightly modified correlator $\tilde{g}^{(2)}_\mathrm{hh}(d)$ defined as
\begin{equation}
	\tilde{g}^{(2)}_{\mathrm{hh}}(\dvec) -1 = \frac{1}{\mathcal{N}_\dvec}\sum_{\ivec} \left(\frac{\langle \hat{n}^h_\ivec \hat{n}^h_{\ivec+\dvec} \rangle}{n^h n^h} - 1\right)
\end{equation}
where compared to Eq.\,\eqref{eq:hh} we replace the normalisation by the \textit{local} densities with the \textit{global} doping level $n^h$. This effectively assumes that the density is homogeneous throughout the system instead of bunched at the centre allowing for easier comparison to the experiment.

\begin{figure}[tb!]
	\centering
	\includegraphics[scale = 1]{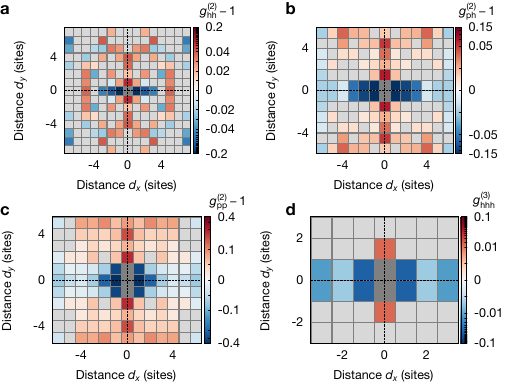}
	\caption{\textbf{Correlator map significance.} Symmetrised hole-hole (\textbf{a}), pair-hole (\textbf{b}), pair-pair (\textbf{c}) and hole-hole-hole (\textbf{d}) correlation maps with errors. All values consistent with zero are set to grey. The signals discussed in the main text are all still clearly visible.}
	\label{fig:maps_err}
\end{figure}

\subsubsection{Statistical significance in correlation maps}\label{sec:map_errs}
The correlation maps shown in Fig.\,\ref{fig:fig2} and \ref{fig:fig3} do not give any indication of which data points in the map are statistically significant or fall below the noise floor of the measurement. To address this, we show in Fig.\,\ref{fig:maps_err} the same maps as in Fig.\,\ref{fig:fig2} and \ref{fig:fig3} where we now set all distances with signals compatible with zero (i.e. the signal being less than 1$\sigma$ away from zero) to grey. All features mentioned in the main text are still clearly visible.

\subsection{Additional coupling terms in the mixD Fermi-Hubbard model}\label{sec:mixD}
In the experiment, we realise a two-dimensional Fermi-Hubbard model with anisotropic tunnel couplings and energy offset on every second site along $y$. In the limit of strong interactions $U\gg t_x,t_y$ used here, this is commonly mapped onto the $t-J$ model. However, this approximation neglects higher-order terms that can arise in the expansion, including a crucial second-order hopping term. While nearest-neighbour hopping is suppressed due to the potential offset, next-nearest neighbour hopping remains resonant in a staggered potential. We experimentally confirmed the presence of this term and its scaling $\tilde{t}_y = t_y'+\frac{t_y^2}{\Delta}$ with direct next-nearest neighbour tunnelling $t'$ (which is however negligible for our parameters)  by performing single-particle quantum walks~\cite{Chalopin2023}. This simple expression neglects interaction effects with atoms in the intermediate lattice site. For $\Delta =0.65(5)\,U$ this means that $\tilde{t}_y \approx \frac{1.54t_y^2}{U}$. This could in principle disfavour stripe formation as the weak Pauli repulsion associated with $\tilde{t}_y$ could inhibit pairs at distance 2 such that only $d_y=1$ hole pairs would form. In this experiment the contribution can mostly be neglected as the principal energy scale is given by $J_y \approx 3 \tilde{t}_y$ which dominates in our parameter regime over $\tilde{t}_y$.

\subsection{Stripe length random data generation}\label{sec:random}
\begin{figure}[t]
	\centering
	\includegraphics[scale = 1]{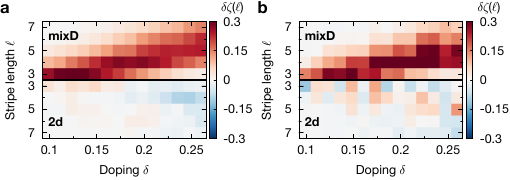}
	\caption{\textbf{Stripe length random data comparison.} \textbf{a}, Comparing experimental data to randomly generated data without any correlations. \textbf{b}, Comparison to randomly placed pairs within the system. Both methods yield qualitatively the same result as the data in the main text.}
	\label{fig:rand}
\end{figure}

To interpret the stripe length results of Fig.\,\ref{fig:fig5}, we compare to random hole distributions with different short-ranged correlations. We first simply randomly sample holes on $9\times9$ sites (see Fig.\,\ref{fig:rand}a) where we observe strong positive signals in the mixD case and negative signals for the 2d case. However, the strong Pauli repulsion along $x$ might have an impact on this signal. For this reason, we randomly sampled holes where we included, in Fig.\,\ref{fig:fig5} of the main text, the experimentally obtained anticorrelations along $x$ (see Fig.\,\ref{fig:fig2}). Finally, we compare to randomly placed pairs along $y$ within the system in Fig.\,\ref{fig:rand}b, exhibiting similar features. Thus we conclude that the observed main qualitative features are relatively insensitive to the exact details of the randomly generated data and that we see a genuine stripe signal that cannot be explained by random or short-ranged correlated holes.
\newpage

\subsection{Numerical simulations of the mixD $\mathbf{t-J}$ model}\label{sec:numerics}
\begin{figure}[tb!]
	\centering
	\includegraphics[scale=1]{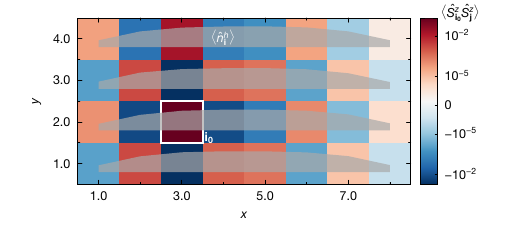}
	\caption{\textbf{Finite temperature DMRG.} DMRG calculations for a $L_x \times L_y = 8 \times 4$ system with periodic boundaries along the short direction, temperature $\kB T/t_x \sim 0.4$, and Hamiltonian parameters as in the experimental setup. Shown are the on-site hole density distributions in each leg, $\langle\hat{n}^h_{\ivec}\rangle$ (grey lines), as well as spin-spin correlations $\langle\hat{S}^z_{\ivec_0} \hat{S}^z_{\jvec}\rangle$ (colour coded) for reference site $\ivec_0 = [x= 3, y=2]$ (white frame). At the maximum hole density distribution in the centre of the chain, a domain wall of the AFM background forms, i.e., a single stripe is observed.}
	\label{fig:ancilla}
\end{figure}

We simulate the mixD $t-J$ model,
\begin{multline}
	\hat{\mathcal{H}} = \hat{\mathcal{P}} \left( -t_x \sum_{\langle\ivec,\jvec\rangle_x} \sum_{\sigma = \uparrow, \downarrow} \hat{c}^{\dagger}_{\ivec,\sigma} \hat{c}_{\jvec,\sigma} + \text{h.c.} \right) \hat{\mathcal{P}} \\
	+ J_x \sum_{\langle\ivec,\jvec\rangle_x} \left( \hat{\mathbf{S}}_{\ivec} \cdot \hat{\mathbf{S}}_{\jvec} - \frac{\hat{n}_{\ivec} \hat{n}_{\jvec}}{4} \right)\\
	+ J_y \sum_{\langle\ivec,\jvec\rangle_y} \left( \hat{\mathbf{S}}_{\ivec} \cdot \hat{\mathbf{S}}_{\jvec} - \frac{\hat{n}_{\ivec} \hat{n}_{\jvec}}{4} \right),
\end{multline} 
(see Eq.~\eqref{eq:tJ} in the main text) for $J_y/t_x=0.5, J_x/J_y=0.3$ at finite temperature using matrix product states (MPS) via mixed state purification schemes~\cite{Feiguin2010, Nocera2016, Paeckel_time}.
In particular, we expand the system by introducing one auxiliary site for each physical site, which allows for displaying mixed physical states as pure states on an enlarged Hilbert space. A pure state in the enlarged system at finite temperature is calculated by evolving the maximally entangled, infinite temperature state $\ket{\Psi(\beta=0)}$ in imaginary time under the physical Hamiltonian, $\ket{\Psi(\tau)} = e^{-\tau \hat{\mathcal{H}}} \ket{\Psi(\beta=0)}$, where $\tau = \beta/2$ with $\beta$ the inverse temperature. Thermal expectation values $\langle\hat{O}\rangle_{T}$ in the physical subset are computed by tracing out the auxiliary degrees of freedom, i.e., 
\begin{equation}
	\langle\hat{O}\rangle_{T} = \frac{\langle\Psi(\beta)|\hat{O}|\Psi(\beta)\rangle}{\langle\Psi(\beta)|\Psi(\beta)\rangle}.
\end{equation}

During the imaginary time evolution, we conserve the particle number in each row $N_{\ell},$ $\ell=1 \dots L_y$, the total particle number in the auxiliary system $N_{\text{aux.}}^{\text{tot}}$, and the total spin $S^{z,\text{tot}}_{\text{phys.+aux.}}$ (the latter allowing for thermal fluctuations of the total magnetization in the physical system). This results in a total of $L_y + 2$ symmetries employed by the DMRG implementation, leading to significant speedups over a single global $\rm{U(1)}$ conservation in the overall physical system~\cite{Schloemer2022}.

The maximally entangled state needed as a starting point of the imaginary time evolution, $\ket{\Psi(\beta=0)}$, is generated using specifically tailored entangler Hamiltonians~\cite{Schloemer2022, Schloemer_HamRecon}. Since these states (being projected product states) are of low bond dimension ($\chi(\tau = 0) \sim \mathcal{O}(100)$), local approximations of the Hamiltonian and subsequent exponentiation will suffer from large projection errors. Hence, we start by employing global methods for a single step in imaginary time, after which the entanglement in the system (and the bond dimension of the thermal MPS) has sufficiently increased to switch to local methods.

Due to the mapping of the (enlarged) 2d system to a 1d chain, the bond dimension required for a fixed accuracy scales exponentially with linear system size in $y-$direction. For doping scans, we limit the system size to $L_x \times L_y = 8\times 3$ with open boundaries, and hole configurations $N_{\ell} = 1,2,3$ for each $\ell = 1,2,3$. For a single hole per chain, we simulate systems up to $L_y = 4$. As this mixD model suffers from the Fermion sign problem, these limited system sizes are still state-of-the-art for numerical calculations while mostly allowing general qualitative comparison to the much larger experimental system.

In particular, we evolve $\ket{\Psi(\beta=0)}$ using global Krylov schemes by a single step $t_x \Delta \tau = 0.01$. Weight cutoffs are set to $10^{-10}$, expanding the bond dimension to $\chi(\tau = \Delta \tau) \sim \mathcal{O}(1000)$. From here on, we switch to the local two-site TDVP method~\cite{Paeckel_time} with time steps of $t_x \Delta \tau = 0.03$, weight and truncation cutoffs of $10^{-10}$ and $10^{-12}$, respectively, and cutting edge maximum bond dimensions of $\chi_{\text{TDVP}} = 30000$. We evolve the system to $\tau t_x = 2.0$, corresponding to a temperature of $\kB T/t_x = 0.25$.
\begin{figure}[tb!]
	\centering
	\includegraphics[scale=1]{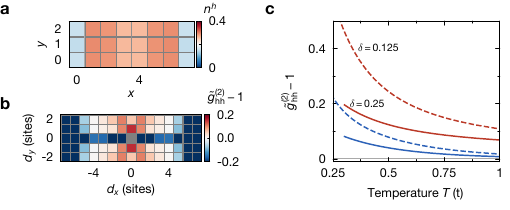}
	\caption{\textbf{Hole correlations in DMRG}. \textbf{a}, Hole density for $L_y=3$, $\delta=0.25$, $\kB T/t_x = 0.41$. Two separate stripes form at this doping level. \textbf{b}, Hole-hole correlations vs distance, reminiscent of the structure shown in Fig.\,\ref{fig:fig2}a. \textbf{c}, Hole correlations as a function of temperature for $d_y=1$ (red) and $d_y=2$ (blue), $\delta=0.125$ (dashed lines) and $\delta=0.25$ (solid lines).}
	\label{fig:dmrg_hh}
\end{figure}
Spin-spin correlations, as well as hole distributions in each leg, are exemplarily shown in Fig.\,\ref{fig:ancilla} for $\kB T/t_x \sim 0.4$ for a system of size $L_x \times L_y = 8 \times 4$ with periodic boundaries along the short direction and $N_{\ell} = 1$ for all $\ell = 1 \dots 4$. At the centre of the chains, where the hole density peaks, an AFM domain wall forms, signalling the formation of a single, fully filled stripe. For a higher doping of $\delta=0.25$, ($L_y=3$, open boundaries), we show the hole density as well as hole-hole correlations in Fig.\,\ref{fig:dmrg_hh}a,b where the two separate stripes are visible. Results as a function of temperature are shown in Fig.\,\ref{fig:dmrg_hh}c for $d_y=1$ and $d_y=2$.

\subsection{Effective descriptions of stripes in the mixD $\mathbf{t-J}$ model}\label{sec:MFT}
\begin{figure}[t]
	\centering
	\includegraphics[scale = 1]{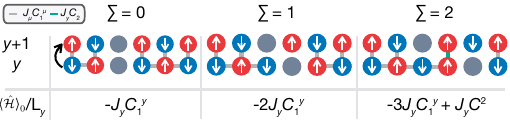}
	\caption{\textbf{Illustration of the effective potential between chain $y$ with its neighbouring chain $y+1$.} Grey lines illustrate energy contributions $\sim J_1^{\mu} C_1^{\mu}$, $\mu=x,y$; green line denotes diagonal correlation with energy contributions $\sim J_y C_2$ starting at $|\Sigma| \geq 2$. Intra-chain energy corrections from the N\'eel state $\sim J_x C_1^x$ are constant and not written down explicitly in the potential.}
	\label{fig:potential}
\end{figure}

\subsubsection{Mean field theory}
In this section, we present a mean field theory (MFT) for the stripe phase in the mixD $t-J$ model. We focus on describing an individual stripe in $y-$direction with exactly one hole per chain, bound by the magnetically mediated confining potentials. In particular, we neglect the interaction between multiple stripes at positions $\ivec_1$ and $\ivec_2$, i.e., we focus on the low doping regime. To illustrate the concept, we first consider a mean field description of the ground state, before generalizing to finite temperature. 

For $t_x\gg J_x,J_y$, quantum correlations between strongly fluctuating holes and spins in squeezed space (defined in \cite{Kruis2004, OgataShiba}) can be neglected~\cite{GrusdtX, Grusdt_tJ, Grusdt_strings, Grusdt_tJz, Bohrdt2020_partons}. Hence, we make the ansatz 
\begin{equation}
	\ket{\psi} = \ket{\psi}_{\text{sq}} \otimes \ket{\psi}_\text{c}, 
\end{equation}
where $\ket{\psi}_{\text{sq}}$ is the spin state of the undoped Heisenberg model in squeezed space, and $\ket{\psi}_{\text{c}}$ is the chargon wave function. Our starting point for the description of the single stripe is the variational Gutzwiller wave function, given by
\begin{equation}
	\ket{\psi}_{\text{c}} = \bigotimes_{y = -\infty}^{\infty} \ket{\phi^{(0)}}_y,
\end{equation}  
i.e., we describe the charge sector by the product of identical single-leg wave functions $\ket{\phi^{(0)}}_y$ in chain $y$. Assuming that the stripe is centred around $x=0$, we express $\ket{\phi^{(0)}}_y$ within the string basis,
\begin{equation}
	\ket{\phi^{(0)}}_y = \sum_{\Sigma = -\infty}^{\infty} \phi^{(0)}_{\Sigma} \ket{y, \Sigma},
\end{equation}
where $\Sigma$ can be understood as the length of the string
measured relative to the centre of the stripe.

Within this variational ansatz, coefficients $ \phi^{(0)}_{\Sigma}$ can be found by minimizing the trial state's energy, $\langle\hat{\mathcal{H}}\rangle_0 = \langle\psi|\hat{\mathcal{H}}|\psi\rangle =  \left(\bra{\psi}_{\text{sq}} \otimes \bra{\psi}_\text{c}\right)  \hat{\mathcal{H}}  \left(\ket{\psi}_{\text{sq}} \otimes \ket{\psi}_\text{c}\right)$,
\begin{multline}
	\frac{\langle\hat{\mathcal{H}}\rangle_0}{L_y} = \frac{E_0}{L_y} - t_x \sum_{\Sigma} \left( \phi^{(0) *}_{\Sigma + 1}  \phi^{(0)}_{\Sigma} + \text{c.c.} \right)\\
	+\sum_{\Sigma, \Sigma'} |\phi^{(0)}_{\Sigma}|^2  |\phi^{(0)}_{\Sigma'}|^2 V_{\text{pot}}(\Sigma - \Sigma').
	\label{eq:var}
\end{multline}
Here, $V_{\text{pot}}(\Sigma)$ is the inter-chain potential defined by the potential energy of two holes in neighbouring chains separated by the string $\Sigma$,
\begin{equation}
	V_{\text{pot}}(\Sigma) = J_y \Big[ \big(|\Sigma| - 1 + \delta_{\Sigma, 0}\big) C_2 - \big(|\Sigma| + 1\big) C_1^y \Big],
	\label{eq:Vpot}
\end{equation}
where $C_1^{\mu} = \langle\psi_{\text{s}} | \hat{\mathbf{S}}_{\ivec} \cdot \hat{\mathbf{S}}_{\ivec+\bb{e}_{\mu}}  |\psi_{\text{s}}\rangle$, $\mu = x,y$, are nearest neighbour, and $C_2 = \langle\psi_{\text{s}} | \hat{\mathbf{S}}_{\ivec} \cdot \hat{\mathbf{S}}_{\ivec+\bb{e}_{x}+\bb{e}_y}  |\psi_{\text{s}}\rangle$ diagonal spin-spin correlations in the undoped Heisenberg model in the ground state. Note that there are also intra-chain contributions, which, however, are constant and only lead to a trivial energy shift on top of the Heisenberg ground state energy $E_0$ (see Fig.~\ref{fig:potential}).

\begin{figure}[t]
	\centering
	\includegraphics[scale = 1]{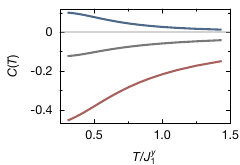}
	\caption{\textbf{Two-point correlators of the squeezed background.} Thermally averaged two-point correlations $C_1^{\mu}(T) = \langle\hat{\mathbf{S}}_{\ivec} \cdot \hat{\mathbf{S}}_{\ivec+\bb{e}_{\mu}}\rangle_T$, $\mu = x$ (grey), $y$ (red) and $C_2(T) = \langle\hat{\mathbf{S}}_{\ivec} \cdot \hat{\mathbf{S}}_{\ivec+\bb{e}_{x}+\bb{e}_y}\rangle_T$ (blue) calculated from DMRG calculations for $J_x/J_y = 0.3$ on a $12 \times 4$ lattice with PBC applied along the short ($y$-) direction.}
	\label{fig:corr_heis}
\end{figure}

By averaging over the upper and lower chain for a given leg, we can reformulate the variational problem, Eq.~\eqref{eq:var}, as a self-consistent ground state search of the mean field Hamiltonian per chain,
\begin{multline}
	\hat{\mathcal{H}}_{\text{MF}} = \frac{E_0}{L_y} - t_x \sum_{\Sigma, \Sigma'} \left[ \hat{h}^{\dagger}_{\Sigma'} \hat{h}_{\Sigma} + \text{h.c.} \right]\\
	+ \sum_{\Sigma} \hat{h}^{\dagger}_{\Sigma} \hat{h}_{\Sigma} V_{\text{eff}}(\Sigma),
	\label{eq:sc_gs}
\end{multline}
where $\hat{h}^{\dagger}_{\Sigma} \ket{y, 0} = \ket{y, \Sigma} $ and 
\begin{equation}
	V_{\text{eff}}(\Sigma) = 2\sum_{\Sigma'} |\phi^{(0)}_{\Sigma'}|^2  V_{\text{pot}} (\Sigma' - \Sigma).
\end{equation}
Note the factor of $2$ in the potential energy, arising from energy contributions between chains $y\pm 1$ with chain $y$. When considering the total energy of the variational wave function, Eq.~\eqref{eq:var}, however, there is no additional factor to not double count inter-chain energy contributions. 

In practice, we set a maximal cutoff for the string length, here chosen as $|\Sigma_{\text{max}}| \sim 15$. By exact diagonalization and self-consistently solving Eq.~\eqref{eq:sc_gs}, the string length distribution $|\phi^{(0)}_{\Sigma}|^2$ within the mean field picture can be calculated. 
\begin{figure}[t]
	\centering
	\includegraphics[scale = 1]{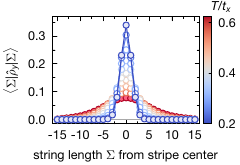}
	\caption{\textbf{String length distribution.} Thermally averaged string length distribution $\langle\Sigma| \hat{\rho}_{\text{MF}}^{(0)} |\Sigma\rangle$ for temperatures $\kB T/t_x = [0.2, 0.625]$, and $t_x/J_y  =2$. Thermal correlations in the Heisenberg model are taken from DMRG calculations with $J_x/J_y = 0.3$ (see Fig.~\ref{fig:corr_heis}).}
	\label{fig:SLD}
\end{figure}

At finite temperature, we generalize the ansatz to a product of density matrices, 
\begin{equation}
	\hat{\rho} = \hat{\rho}_{\text{sq}}\otimes \left( \bigotimes_{y = -\infty}^{\infty} \hat{\rho}_{\text{MF}}^{(0)}\right),
\end{equation} 
where 
\begin{equation}
	\hat{\rho}_{\text{MF}}^{(0)} = \frac{1}{Z} e^{-\beta \hat{\mathcal{H}}_{\text{MF}}(\hat{\rho}_{\text{MF}}^{(0)}, T)}
\end{equation} 
defines the self-consistency equation via
\begin{equation}
	\begin{gathered}
		\frac{\hat{\mathcal{H}}_{\text{MF}}(\hat{\rho}_{\text{MF}}^{(0)})}{L_y} = \frac{E_0}{L_y} - t_x \sum_{\Sigma, \Sigma'} \left[ \hat{h}^{\dagger}_{\Sigma'} \hat{h}_{\Sigma} + \text{h.c.} \right]\\
		+ \sum_{\Sigma} \hat{h}^{\dagger}_{\Sigma} \hat{h}_{\Sigma} V_{\text{eff}}(\Sigma; \hat{\rho}_{\text{MF}}^{(0)},  T), \\
		V_{\text{eff}}(\Sigma; \hat{\rho}_{\text{MF}}^{(0)}, T) = \\
		2\sum_{\Sigma'} \langle\Sigma| \hat{\rho}_{\text{MF}}^{(0)} |\Sigma\rangle  V_{\text{pot}} (\Sigma' - \Sigma; T).
	\end{gathered}
\end{equation}
Here, $C_1^{\mu}(T) = \langle\hat{\mathbf{S}}_{\ivec} \cdot \hat{\mathbf{S}}_{\ivec+\bb{e}_{\mu}}\rangle_T$, $\mu = x,y$ and $C_2(T) = \langle\hat{\mathbf{S}}_{\ivec} \cdot \hat{\mathbf{S}}_{\ivec+\bb{e}_{x}+\bb{e}_y}\rangle_T$ entering $V_{\text{pot}}$ in Eq.~\eqref{eq:Vpot} are thermally averaged two-point correlators of the 2d Heisenberg model. Given the self-consistent solution of $\hat{\rho}_{\text{MF}}^{(0)}$, the mean field string length distribution is determined by the diagonal elements of $\hat{\rho}_{\text{MF}}^{(0)}$, i.e., $p_{\Sigma} = \langle\Sigma| \hat{\rho}_{\text{MF}}^{(0)} |\Sigma\rangle $.

We use finite temperature DMRG methods (see \hyperref[sec:numerics]{previous section}) to calculate thermally averaged nearest-neighbour and diagonal correlations of the undoped Heisenberg model with $J_x/J_y = 0.3$ on a $L_x \times L_y = 12 \times 4$ lattice with periodic boundaries along $y$, see Fig.~\ref{fig:corr_heis}. Results for the corresponding mean field estimates of the string length distributions in the stripe phase are shown in Fig.~\ref{fig:SLD} for $t_x/J_y = 2$ and temperatures $\kB T/t_x = [0.2, 0.625]$.

\begin{figure}[t!]
	\centering
	\includegraphics[scale = 1]{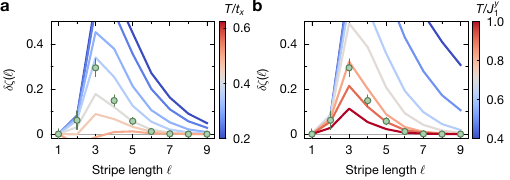}
	\caption{\textbf{Stripe length distributions for effective theories.} \textbf{a}, Difference of stripe lengths from MFT to random distribution for temperatures of $\kB T/t_x \in [0.2,0.5]$ and $J_x/J_y=0.3$ and experimental data for $\delta = 0.083$ (markers) as in the inset of Fig.\,\ref{fig:fig5}a of the main text. \textbf{b}, Stripe length histograms using the classical MHZ estimate for temperatures of $\kB T/J_1^y \in [0.4,1]$ which shows qualitatively similar results. }
	\label{fig:MFT_MHZ_exp}
\end{figure}

Using the MFT string length distributions, we sample snapshots and compare the resulting stripe length distributions to the experiment (see Fig.~\ref{fig:MFT_MHZ_exp}a). At the expected temperature of $\kB T/t_x \approx 0.3$, the effective description matches the experiment rather well with only a slight overestimation of the order in the mean field description.

\subsubsection{M\"uller-Hartmann-Zittartz estimate}
To make further comparisons to statistical models, we reduce the mixD system to a 1d, purely classical model of fluctuating holes bound together by the effective potential $V_{\text{pot}}$, Eq.~\eqref{eq:Vpot} (M\"uller-Hartmann-Zittartz (MHZ) approach). Denoting with $x_{\ell}$ the $x-$position of the doped hole in leg $\ell$ (we again consider one hole per chain, i.e. a single fluctuating domain wall), the effective Hamiltonian (excluding quantum fluctuations from the hopping of the holes) for a system of size $(L_x+1) \times (L_y + 1)$ reads
\begin{equation}
	\hat{\mathcal{H}}_{\text{MHZ}} = \sum_{\ell=1}^{L_y} V_{\text{pot}}(|x_{\ell} - x_{\ell +1}|; T), 
\end{equation}
where again the temperature dependent correlators $C_1^{\mu}(T) = \langle\hat{\mathbf{S}}_{\ivec} \cdot \hat{\mathbf{S}}_{\ivec+\bb{e}_{\mu}}\rangle_T$, $\mu = x,y$ and $C_2(T) = \langle\hat{\mathbf{S}}_{\ivec} \cdot \hat{\mathbf{S}}_{\ivec+\bb{e}_{x}+\bb{e}_y}\rangle_T$ enter the effective potential $V_{\text{pot}}(|x_{\ell} - x_{\ell +1}|; T)$ in Eq.~\eqref{eq:Vpot}.

\begin{figure}[t!]
	\centering
	\includegraphics[scale=1]{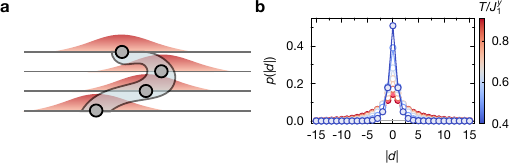}
	\caption{\textbf{M\"uller-Hartmann-Zittartz calculations.} \textbf{a}, Illustration of the MHZ estimate. Hole positions are sampled according to the red probability distributions, whereby the distribution for position $x_{\ell+1}$ is centred around $x_{\ell}$ -- capturing fluctuating, extended stripes. \textbf{b}, Hole distance distributions, Eq.~\eqref{eq:pd}, for various temperatures $\kB T/J_1^y = 0.4 \dots 0.9$. }
	\label{fig:MHZ}
\end{figure}

The partition function, $Z$, decouples when being expressed solely by distances $d_{\ell} = x_{\ell} - x_{\ell + 1}$,
\begin{equation}
	\begin{aligned}
		Z &= \sum_{\{x_{\ell}\}} \prod_{\ell = 1}^{L_y} \text{exp}\left[-\beta V_{\text{pot}}(|x_{\ell} - x_{\ell +1}|; T)\right] \\ 
		&= \sum_{d_1 = -L_x}^{L_x} \dots \sum_{d_{L_y} = -L_x}^{L_x} \prod_{\ell = 1}^{L_y} \text{exp}\left[-\beta V_{\text{pot}}(|d_{\ell}|; T)\right] \\
		& = \left[\sum_{d = -L_x}^{L_x}  \text{exp}\left[-\beta V_{\text{pot}}(|d|; T)\right] \right]^{L_y} = [Z_1]^{L_y}.
	\end{aligned}
\end{equation}
The probability of finding two adjacent holes at distance $d$ in chains $\ell, \ell + 1$ is given by
\begin{equation}
	p(d) = \text{exp}\left[ -\beta V_{\text{pot}}(|d|; T) \right]/Z_1,
	\label{eq:pd}
\end{equation}
shown for various temperatures $\kB T/J_1^y$ in Fig.~\ref{fig:MHZ}b.

Fixing the first hole in the centre and sampling distances according to Eq.~\eqref{eq:pd}, we again generate snapshots of the hole configurations. Note that, while in the MFT fluctuating stripes pinned to the centre were described, the classical formulation as given above captures stripes that are not pinned to the boundary, and hence naturally form extended hole configurations as illustrated in Fig.~\ref{fig:MHZ}a (see also Fig.\,\ref{fig:fig5}a in the main text). We compare the results to the experimental data in Fig.\,\ref{fig:MFT_MHZ_exp}b where for $\kB T/J_1^y =0.8$ we observe similar features to the experiment and the results from MFT for $\kB T/t_x=0.36$.

\bigskip
\end{document}